# Multi-rotational switching in a noncollinear antiferromagnet by spin-orbit torque

Yuma Sato[1,2], Yutaro Takeuchi[3,4*], Yuta Yamane[1,5*], Shun Kanai[1,2,4,6,7,8], and Shunsuke Fukami [1,2,4,8,9,10*]

[1]*Laboratory for Nanoelectronics and Spintronics, Research Institute of Electrical Communication, Tohoku University, Sendai 980-8577, Japan*

[2]*Department of Electronic Engineering, Graduate School of Engineering, Tohoku University, Sendai 980-8579, Japan*

[3]*Research Center for Magnetic and Spintronic Materials, National Institute for Materials Science, Tsukuba 305-0047, Japan*

[4]*WPI-Advanced Institute for Materials Research, Tohoku University, Sendai 980-8577, Japan*

[5]*Frontier Research Institute for Interdisciplinary Sciences, Tohoku University, Sendai 980-8578, Japan*

[6]*Division for the Establishment of Frontier Sciences of Organization for Advanced Studies, Tohoku University, Sendai 980-8577, Japan*

[7]*National Institutes for Quantum Science and Technology, Takasaki 370-1292, Japan*

[8]*Center for Science and Innovation in Spintronics, Tohoku University, Sendai 980-8577, Japan*

[9]*Center for Innovative Integrated Electronic Systems, Tohoku University, Sendai 980-8572, Japan*

[10]*Inamori Research Institute for Science, Kyoto 600-8411, Japan*

*e-mail: TAKEUCHI.Yutaro@nims.go.jp, yuta.yamane.e8@tohoku.ac.jp, s-fukami@riec.tohoku.ac.jp

## Abstract

Spintronics has advanced through discoveries of various electrically-driven spin dynamics in nanomagnets. Here, we report a novel switching dynamics of spin systems driven by spin-orbit torque, using a noncollinear antiferromagnetic nanodot. With electric pulses spanning a wide range of durations and amplitudes, we find an unconventional insensitivity of a threshold current density to pulse duration in switch-back events. This observation is attributed to a previously unrecognized process, in which the noncollinear antiferromagnetic order undergoes multiple rotations before completing reversal – a phenomenon we term multi-rotational switching. Our theoretical analysis reveals that multi-rotational switching arises from the interplay of three key factors: current-driven coherent rotation of the noncollinear antiferromagnetic order, field-induced reorientation of the uncompensated net magnetization, and thermal fluctuations. These findings establish a microscopic mechanism governing current-induced switching in noncollinear antiferromagnets, a topic of growing interest for next-generation spintronics technologies, opening a new route to controlling antiferromagnetic order in nanodevices.

## Introduction

Antiferromagnets (AFMs) exhibit rich and fascinating spin physics, and are expected to play important roles in next-generation information technologies[1–4]. Their spin structures are inherently robust against external magnetic disturbances owing to their vanishingly small net magnetization. Moreover, their inertial spin dynamics and exchange-enhanced frequencies enable device operations that are faster and more energy efficient than those of conventional magnetic systems[5–8]. Among various classes of AFMs, noncollinear AFMs represented by $Mn_3X$ ($X$ = Sn, Ge and so on) have attracted particular interest in spintronics. They are characterized by a triangular spin order, that is, a noncollinear AFM order, realized on the stacked Kagome lattice formed by Mn ions. Noncollinear AFMs not only enjoy the general advantages of AFMs, but also host a variety of exotic phenomena associated with their topological electronic band structures and the magnetic symmetry of the noncollinear AFM order[9–21].

Manipulation of AFM order is a key requirement for AFM-based technologies. Recent studies have demonstrated that noncollinear AFM order can be driven into continuous rotational motion by all-electrical means[7,22,23], exploiting spin-orbit torque (SOT)[24–27]. This phenomenon is understood based on angular-momentum transfer from injected spin currents to the rotational dynamics of the noncollinear AFM order[28–30]. Deterministic bipolar switching of noncollinear AFM order utilizing SOT has also been reported with the assistance of an in-plane magnetic field, where the switching polarity is set by the directions of the current and the field[31–36]. This SOT-switching scheme superficially resembles its ferromagnetic counterpart, although the equation of motion governing noncollinear AFM order differs fundamentally from that of ferromagnetic magnetization[7,28,30,35,37]. Several mechanisms have been proposed to explain this bipolar SOT switching in both epitaxial[7,22,32,35] and polycrystalline[31,33,34,38] thin films. For epitaxial films, which are particularly important for uncovering the fundamental aspects of noncollinear AFM dynamics, the observed switching has been attributed to a competition between current-driven rotation of the noncollinear AFM order and field-induced reorientation of the uncompensated net magnetization[7,22,32,35]. Despite these extensive studies, the microscopic dynamics of noncollinear AFM order underlying the bipolar SOT switching remain incompletely understood, limiting further exploration of the functional potential of noncollinear AFMs.

Here we reveal a previously unrecognized dynamical mode of noncollinear AFM order that emerges in bipolar SOT switching. Using a noncollinear AFM nanodot, we observe an unconventional switching phase diagram as a function of current-pulse duration and amplitude, in which the switching-back threshold current density depends only weakly on the pulse duration. We show that this is explained by the spatial asymmetry introduced by the in-plane field, combined with SOT and thermal fluctuations, causing the noncollinear AFM order to undergo multiple rotations before completing the switching. We refer to this phenomenon as multi-rotational switching. The experimental observations are well captured by a theoretical model we develop, which reproduces the key features of the switching phase diagram. Our finding of multi-rotational switching sheds new light on the microscopic process underlying the SOT switching of AFM order, and opens a pathway to novel device concepts enabled by AFM-based SOT, including robust switching over a wide range of pulse durations, from nanoseconds to milliseconds and beyond.

## Sample fabrication and characterization

To investigate single-domain noncollinear AFM dynamics, we fabricate a heterostructure incorporating a nanodot of $Mn_3Sn$ (Fig. **1a**). The stack structure consists of W(2 nm)/Ta(3 nm)/$Mn_3Sn$(20 nm)/MgO(1.3 nm)/Ru(1 nm), deposited on a MgO (110) substrate by magnetron sputtering. The epitaxial $Mn_3Sn$ film exhibits $(1\bar{1}00)$-plane, or M-plane, crystallographic orientation[22,35,39], where the Kagome planes in $Mn_3Sn$ are perpendicular to the film plane. The stack is processed into a Hall device with a $Mn_3Sn$ nanodot formed on a cross-shaped W/Ta channel[7,40]. The nominal dot diameter is 200 nm, and the channel width and length are 250 nm and 600 nm, respectively. Displayed in Fig. **1b** is an image of the fabricated nanodot sample, captured by scanning electron microscopy. See Materials and methods for more details about sample preparation.

To confirm the noncollinear AFM order in our $Mn_3Sn$ nanodot, we measure the anomalous Hall resistance $R_{\mathrm{H}}$ as a function of the out-of-plane field $H_z$ (Fig. **1c**). We then establish the SOT-driven rotation and the field-assisted bipolar switching of the noncollinear AFM order in the DC regime. Figure **1d** shows $R_{\mathrm{H}}$ as a function of $J_{\mathrm{HM}}$, the current density in the heavy metal (W/Ta) layers, under various in-plane fields $H_x$ applied parallel to the current. With $H_x = 0$, the fluctuation in $R_{\mathrm{H}}$ is observed above sufficiently large $J_{\mathrm{HM}}$, arising from the SOT-driven continuous rotation of the noncollinear AFM order[22]. When a nonzero $H_x$ is applied, bipolar switching of $R_{\mathrm{H}}$ is observed. The threshold current density decreases with increasing $H_x$, and the switching polarity depends on the sign of $H_x$. This switching behaviour can be attributed to the SOT-driven rotation of the noncollinear AFM order being balanced by $H_x$, which favors alignment of the uncompensated net magnetization along its direction[32].

## Switching probability measurements

Next, we perform switching experiments using current pulses with pulse duration $\tau_P$ ranging from 5 ns to 5 ms, along with a DC in-plane magnetic field (Fig. **1a**). Before each pulse, the $Mn_3Sn$ nanodot is initialized to the "down" state, where $R_H < 0$. The nanodot is considered "switched" when the post-pulse state is "up" ($R_H > 0$). See Materials and methods and Supplementary Figs. **1** and **2** for more details on the experimental procedure.

Figures **2a-d** show the switching probability $P$ as a function of $J_{HM}$, under $\mu_0 H_x = 100$ mT and for four different $\tau_P$. Here, $P$ is defined as the ratio of successful switching events out of 60 pulse applications and $\mu_0$ is the vacuum permeability. A common trend across all four cases is that $P$ initially increases with $J_{HM}$ and reaches unity. With further increase in $J_{HM}$, $P$ starts to decrease and asymptotically approaches 0.5. We define the critical current density $J_{C1}$ ($J_{C2}$) as the current density at which $P$ first touches 0.5 (0.75) during the switching (switching-back) process for a given $\tau_P$. The asymmetry between the critical currents for the switching and switching-back arises from the broken $\pm x$ symmetry due to $H_x$. In the present configuration, $H_x$ assists the down-to-up rotation of the noncollinear AFM order (switching) while it opposes the up-to-down (switching-back)[41].

The dependence of $P$ on $(J_{HM}, \tau_P)$ for the case of $\mu_0 H_x = 100$ mT is summarized by the two-dimensional color map in Fig. **2e**. (The magnetic-field dependence of the switching properties will be discussed later.) It is observed that $J_{C1}$ decreases with increasing $\tau_P$, whereas $J_{C2}$ is largely insensitive to $\tau_P$. The $\tau_P$ dependence of $J_{C1}$ can be readily understood: with thermal fluctuations, a longer-duration pulse increases the probability of thermally-assisted switching. A similar pulse-duration dependence of the switching current has also been reported in ferromagnets[26,42]. In contrast, the weak $\tau_P$ dependence of $J_{C2}$ defies such a straightforward explanation, with no comparable phenomenon known in ferromagnets. In what follows, we discuss the microscopic origin of this unconventional switching behaviour.

## Numerical simulation and theoretical modelling

We model the $Mn_3Sn$ nanodot by a three-sublattice AFM with each sublattice ($A$, $B$, and $C$) carrying the same saturation magnetization $M_S$. The sublattice magnetizations $\boldsymbol{m}_A$, $\boldsymbol{m}_B$, and $\boldsymbol{m}_C$ are assumed to obey the coupled Landau–Lifshitz–Gilbert (LLG) equations[43], with thermal fluctuations incorporated according to the fluctuation-dissipation theorem[44]. In general, the noncollinear AFM order formed by the sublattice magnetizations in $Mn_3Sn$ can be described by two Néel vectors[37], or equivalently a cluster octupole moment[45]. Since all sublattice magnetizations are well confined in the Kagome planes, however, the noncollinear AFM order can be represented by a single Néel vector $\boldsymbol{n} = \frac{\boldsymbol{m}_A + \boldsymbol{m}_B - 2\boldsymbol{m}_C}{3\sqrt{2}}$ [28,30]. The three sublattice magnetizations maintain nearly 120° orientations with each other, ensuring that $\sqrt{2}|\boldsymbol{n}| \simeq 1$. See Materials and methods for further details of our model and calculations. Figure **2f** shows the calculated switching probability $P(J_{\mathrm{HM}}, \tau_{\mathrm{P}})$ from the down ($\sqrt{2}n_z \simeq -1$) to up ($\sqrt{2}n_z \simeq +1$) state, obtained by numerically solving the coupled LLG equations with $\mu_0 H_x = 100\,\mathrm{mT}$ and $\tau_{\mathrm{P}}$ ranging from 2 ns to 100 ns. Here, $P$ is defined in the same way as in Fig. **2e**. The simulation successfully reproduces the experimentally observed trends of $J_{\mathrm{C1}}$ and $J_{\mathrm{C2}}$ as functions of $(J_{\mathrm{HM}}, \tau_{\mathrm{P}})$.

To understand the origin of the weak $\tau_{\mathrm{P}}$ dependence of $J_{\mathrm{C2}}$, we now develop a theoretical model of the Néel vector dynamics. Figure **2g** summarizes the key results of our analysis, the derivation of which is presented in the following. Specifically, the switched region between $J_{\mathrm{C1}}$ and $J_{\mathrm{C2}}$ (the red region in Fig. **2f**) will be shown to consist of two sub-regimes (Fig. **2g**), separated by an additional threshold current density $J_{\mathrm{multi}}$ defined below. It will become clear that, in the region between $J_{\mathrm{multi}}$ and $J_{\mathrm{C2}}$, the Néel vector undergoes multiple rotations before completing the reversal, that is, multi-rotational switching, and that the observed insensitivity of $J_{\mathrm{C2}}$ to $\tau_{\mathrm{P}}$ is rooted in the nature of this multi-rotational switching. These findings convey a central message of this work.

We introduce an angular variable $\theta$ as in the inset of Fig. **3a**. The coupled LLG equations for the three sublattice magnetizations are reduced to the equation of motion for $\theta$ [7,28,30]

$$\frac{1}{\gamma H_{\mathrm{E}}} \frac{\partial^2 \theta}{\partial t^2} = -\alpha \frac{\partial \theta}{\partial t} - \frac{\gamma H_K}{2} \sin 2\theta - \gamma H_x \frac{M_{UC}}{3M_S} \cos\theta - \gamma D_{\mathrm{S}} J_{\mathrm{HM}} + F_{\mathrm{th}}(T). \qquad (1)$$

We outline a derivation of Eq. (1) in Materials and methods. Here, the left-hand side represents the inertial term, with $\gamma$ denoting the gyromagnetic ratio and $H_{\mathrm{E}}$ the effective field associated with the antiferromagnetic exchange coupling. The right-hand side contains five "forces" originating from: (1) the damping torques with $\alpha$ the Gilbert damping constant, (2) the global uniaxial anisotropy characterized by the effective field $H_K$, (3) Larmor torques from the external field $H_x$ with $M_{\mathrm{UC}}$ the uncompensated net magnetization, (4) the SOT with $D_{\mathrm{S}}$ a constant describing the Slonczewski-like torque, and (5) thermal fluctuations with $F_{\mathrm{th}}(T)$ the Langevin force and $T$ the sample temperature. The Néel vector dynamics is, therefore, mathematically equivalent to that of a point mass $m = (\gamma H_{\mathrm{E}})^{-1}$ subject to the frictional $(\alpha)$, stochastic, and conservative forces, with the latter arising from a tilted-washboard potential

$$u(\theta) = \frac{\gamma H_K}{2}\sin^2\theta + \gamma H_x \frac{M_{UC}}{3M_S}\sin\theta + \gamma D_{\mathrm{S}} J_{\mathrm{HM}}\theta, \quad (2)$$

given in the unit of angular frequency. Under the condition $\frac{\gamma H_K}{2} \gg \gamma H_x \frac{M_{UC}}{3M_S}$, the first two terms stabilize the up state $(\theta = 2n\pi)$ and the down state $[\theta = (2n-1)\pi]$ by creating asymmetric potential barriers between them (Fig. **3a**), where $n$ is an integer. With $H_x > 0$, the barrier $\Delta_{\mathrm{down}\to\mathrm{up}}$, for the "right-going" down-to-up transition, is lower than $\Delta_{\mathrm{up}\to\mathrm{down}}$, for the "right-going" up-to-down. The last term in Eq. (2), linear in $\theta$, introduces a global tilt to the potential. Notice that $D_{\mathrm{S}} < 0$ for our system due to the negative sign of the effective spin Hall angle of Ta/W underlayer (see Materials and methods). Both $\Delta_{\mathrm{down}\to\mathrm{up}}$ and $\Delta_{\mathrm{up}\to\mathrm{down}}$ decrease with increasing $J_{\mathrm{HM}}(\geq 0)$, and vanish at $J_{\mathrm{HM}} \approx J_{\mathrm{C1}}^* \equiv \frac{1}{|D_{\mathrm{S}}|}\left(\frac{1}{2}H_K - \frac{M_{UC}}{3\sqrt{2}M_S}H_x\right)$ and $J_{\mathrm{C2}}^* \equiv \frac{1}{|D_{\mathrm{S}}|}\left(\frac{1}{2}H_K + \frac{M_{UC}}{3\sqrt{2}M_S}H_x\right)$, respectively.

In the framework of Eq. (2), $J_{\mathrm{C1}}$ corresponds to the current density at which $\Delta_{\mathrm{down}\to\mathrm{up}}$ is sufficiently reduced for the down-to-up transition to become thermally activated, causing $P$ to reach 0.5, while $\Delta_{\mathrm{up}\to\mathrm{down}}$ remains too large to be overcome. This is schematically illustrated in Fig. **3a**. Starting from the initial state $\theta = \pi$, the octupole moment passes through the barrier $\Delta_{\mathrm{down}\to\mathrm{up}}$ with thermal assistance and settles in the up state $\theta = 2\pi$, completing a $\pi$ switching. Figure **3b** displays a result of the LLG simulation, showcasing a representative example of the Néel vector dynamics corresponding to Fig. **3a**. The plot shows the time evolutions of the z component of the Néel vector (normalized to unity) for $\tau_{\mathrm{P}} = 10$ ns and $\mu_0 H_x = 100$ mT. At zero temperature, $J_{\mathrm{C1}}$ coincides with $J_{\mathrm{C1}}^*$.

As $J_{\mathrm{HM}}$ is further increased beyond $J_{\mathrm{C1}}^{*}$ such that $\Delta_{\mathrm{up}\to\mathrm{down}}$ is sufficiently lowered, the down-to-up transition becomes dominated by coherent SOT driving, whereas the up-to-down transition is thermally activated (Fig. **3c**): The Néel vector therefore experiences multiple rotations during a pulse. Once a thermally-assisted up-to-down transition occurs (e.g., $2\pi$-to-$3\pi$), the Néel vector is immediately driven back to the up state ($4\pi$) by SOT. Consequently, the Néel vector predominantly resides in the up state during the pulse, resulting in a high probability of being found in the up state after the pulse. A result of the LLG simulation corresponding to Fig. **3c** is shown in Fig. **3d**.

We define $J_{\mathrm{multi}}$ as the current density at which the probability $P_{\pi}$ of a single $\pi$ rotation drops to 0.5 due to the onset of multiple rotations. (Switching events involving $\pi$, $3\pi$, $5\pi$, … rotations are all counted in $P$, whereas only the $\pi$ rotation contributes to $P_{\pi}$.) The regime $J_{\mathrm{multi}} \leq J_{\mathrm{HM}} < J_{\mathrm{C2}}$ is thus identified as the multi-rotational switching regime (Fig. **2g**), where $P \cong 1$ while $P_{\pi} \ll 1$ over most of this regime. Accordingly, the switched regime $(J_{\mathrm{C1}} \leq J_{\mathrm{HM}} < J_{\mathrm{C2}})$ consists of two distinct sub-regimes: the $\pi$-switching regime $(J_{\mathrm{C1}} \leq J_{\mathrm{HM}} < J_{\mathrm{multi}}$ where $P_{\pi} \cong 1)$ and the multi-rotational switching regime $(J_{\mathrm{multi}} \leq J_{\mathrm{HM}} < J_{\mathrm{C2}}$ where $P_{\pi} \ll 1)$. The former follows a thermally-assisted mechanism analogous to that in ferromagnets, whereas the latter reveals a distinctive dynamical mode characteristic of AFMs.

Once $J_{\mathrm{HM}}$ exceeds $J_{\mathrm{C2}}^{*}$, both potential barriers disappear, and $\theta$ undergoes continuous downhill motion (Fig. **3e**). The corresponding Néel vector dynamics is a continuous coherent rotation primarily driven by SOT (Fig. **3f**). The numerically obtained $J_{\mathrm{C2}}$ is well reproduced by the analytical $J_{\mathrm{C2}}^{*}$(Fig. **2g**). Establishing a rigorous theoretical connection between $J_{\mathrm{C2}}^{*}$ and $J_{\mathrm{C2}}$ is nontrivial, since the latter is defined through the switching probability at finite temperature. Nonetheless, the close quantitative agreement between $J_{\mathrm{C2}}^{*}$ and $J_{\mathrm{C2}}$, for the parameter values relevant to a $Mn_3Sn$ nanodot, strongly indicates that the weak $\tau_{\mathrm{P}}$ dependence of $J_{\mathrm{C2}}$ originates from the transition between multi-rotational switching and coherent rotation: Once the potential barriers vanish, the Néel vector rotates freely without thermal activation, making the switching threshold essentially independent of pulse duration.

We now revisit the numerical result for the switching probability $P(J_{\mathrm{HM}}, \tau_{\mathrm{P}})$ (Fig. **2f**), and decompose it into the probabilities $P_{(2n-1)\pi}$ of $(2n-1)\pi$ rotations. Figure **4a** shows $P_{\pi}$ as a function of $(J_{\mathrm{HM}}, \tau_{\mathrm{P}})$, where $P_{\pi} \cong 1$ within the $\pi$-switching regime $(J_{\mathrm{C1}} \leq J_{\mathrm{HM}} < J_{\mathrm{multi}})$. Correspondingly, $P_{(2n-1)\pi}$ for $n > 1$ attain appreciable values for $J_{\mathrm{multi}} \leq J_{\mathrm{HM}}$ (Figs. **4b** and **c**). The summation $\sum_{n>1} P_{(2n-1)\pi}$ of the switching probabilities with multiple rotations is plotted in Fig. **4d**: It reaches unity in the regime $J_{\text{multi}} \leq J_{\text{HM}} < J_{\text{C2}}$, confirming that the switching in this regime takes place overwhelmingly via the multi-rotational switching. In the coherent-rotation regime $(J_{\text{C2}} \leq J_{\text{HM}})$, $\sum_{n>1} P_{(2n-1)\pi}$ asymptotically approaches 0.5 from above as $J_{\text{HM}}$ increases. This is because Néel vector dynamics is so rapid (Fig. **3f**) that the final state, up or down, becomes effectively stochastic, as the thermal fluctuation spoils the time coherence in the Néel vector dynamics. Perfect coherence is restored at zero-temperature for $J_{\text{C2}} \leq J_{\text{HM}}$[7]. The results in Fig. **4** clearly demonstrate, once again, that for $J_{\text{C1}} \leq J_{\text{HM}}$, Néel vector dynamics under SOT and an in-plane field falls into the three distinct regimes.

## Magnetic-field dependence of switching properties

Finally, we investigate the magnetic-field dependence of the switching behaviour. Figures **5a-c** show the experimental $P(J_{\text{HM}}, \tau_{\text{P}})$ for three different values of $H_x$. As $H_x$ increases, $J_{\text{C1}}$ and $J_{\text{C2}}$ are shifted overall downwards and upwards, respectively. This is understood as $\Delta_{\text{down}\to\text{up}}$ $(\Delta_{\text{up}\to\text{down}})$ is lowered (raised) by $H_x$, which is consistent with what is predicted from Eq. (2). This trend is well reproduced by the LLG simulations (Figs. **5d-f**). In Fig. **5g**, the open circles represent the experimental $\overline{J_{\text{C2}}}(H_x)$, defined as the $\tau_{\text{P}}$-averaged $J_{\text{C2}}(\tau_{\text{P}}, H_x)$, with the error bars indicating the standard deviation. The numerically obtained $\overline{J_{\text{C2}}}(H_x)$ (open squares with error bars) and the theoretical $J_{\text{C2}}^*$ (solid line) are also shown. The close quantitative agreement among these three datasets provides strong additional support for the validity of our theoretical model. No switched regime is experimentally observed under zero magnetic field (Materials and methods and Supplementary Fig. **3**), confirming that both the $\pi$ switching and multi-rotational switching require the asymmetric potential barriers arising from the in-plane field.

## Summary

We have revealed multi-rotational switching, a novel dynamical mode at play in the SOT switching of noncollinear AFM order, in a nanodot device. The unconventional switching behaviour experimentally observed is well reproduced by numerical simulations, and the underlying physics is captured by our theoretical modelling of Néel vector dynamics under an effective potential energy landscape. Since the factors necessary for multi-rotational switching, namely SOT-driven rotation of the AFM order, field-induced reorientation of the magnetization and thermal fluctuations, are universally expected in easy-plane AFMs, the discovered mechanism is broadly relevant to a wide range of spintronic systems of current interest.

The multi-rotational switching reported in this work enables reliable operation of SOT devices across a broad area in the $(J_{\mathrm{HM}}, \tau_{\mathrm{P}})$ space, owing to the thermal robustness of $J_{\mathrm{C2}}$ and its tunability via an external magnetic field. The insensitivity of $J_{\mathrm{C2}}$ to thermal disturbance, and thus to $\tau_{\mathrm{P}}$, also points to the potential for a single AFM device to function as both a memory element and an oscillator. This dual functionality can be accessed by simply tuning $J_{\mathrm{HM}}$ below $J_{\mathrm{C2}}$ (memory mode) or above $J_{\mathrm{C2}}$ (oscillator mode). The thermal robustness of $J_{\mathrm{C2}}$ is further demonstrated by numerical simulations, by confirming its insensitivity to sample size, as shown in Materials and methods and Supplementary Fig. **4**.

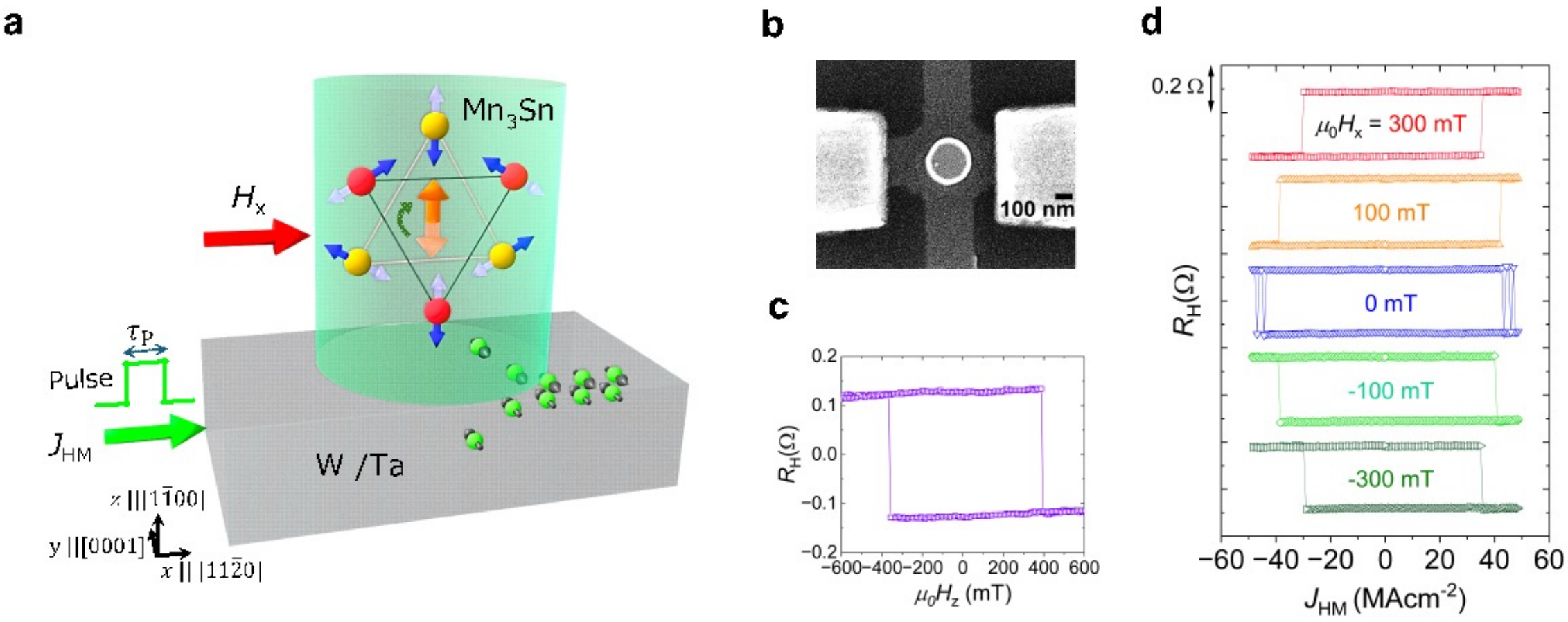

**Fig. 1. Schematic and characterizations of the Mn3Sn nanodot sample. a**, SOT switching of the noncollinear AFM order in the Mn3Sn nanodot, with the orange arrow indicating the Néel vector. **b**, Scanning electron microscope image of the fabricated Hall device. **c**, $R_H$-$H_z$ curve for the Mn3Sn nanodot. **d**, $R_H$-$J_{\mathrm{HM}}$ curves in DC experiments (the current pulse duration being 100 ms) with various in-plane magnetic field $H_x$, where $\mu_0$ is the vacuum permeability.

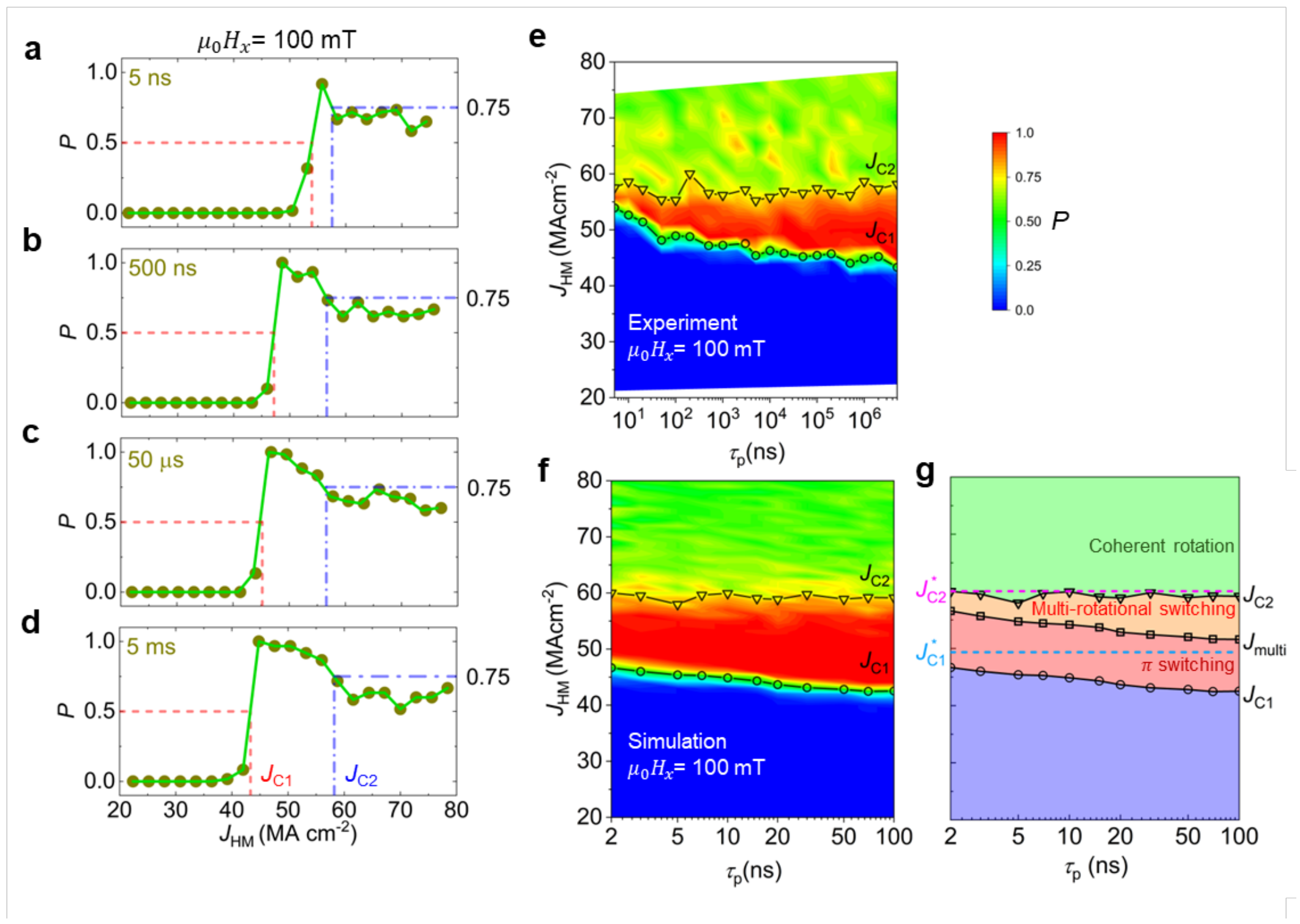

**Fig. 2. Switching probability measurements. a-d,** $P$ versus $J_{\mathrm{HM}}$ with four different $\tau_P$. **e-f,** Two-dimensional mappings of $P(J_{\mathrm{HM}}, \tau_P)$ with $\mu_0 H_x = 100$ mT, obtained by experiment (**e**) and simulation (**f**). **g,** Phase diagram identifying the three distinct switching regimes for the numerical result in **f**.

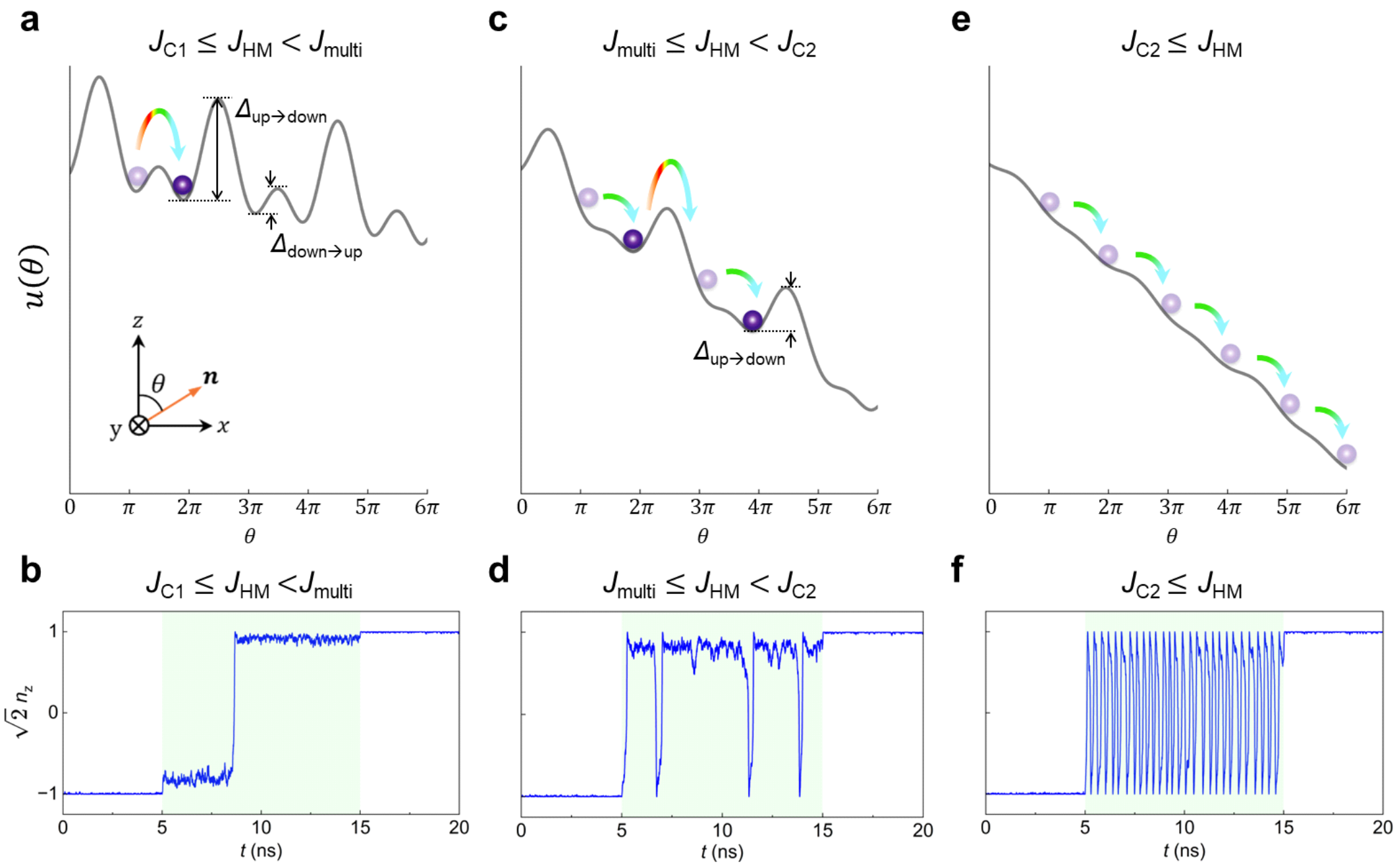

**Fig. 3. Theoretical and numerical analyses on the SOT switching of the Néel vector.** Schematic illustrations of Néel vector dynamics based on the analytical model with Eqs. (1) and (2) in the $\pi$-switching regime (**a**), the multi-rotational switching regime (**c**), and the coherent rotation regime (**e**). The unit of $u(\theta)$ is arbitrary. Numerical simulations for the time evolution of the Néel vector in the corresponding regimes are shown in (**b**), (**d**) and (**f**). In the simulation, the current pulse is applied from 5 ns to 15 ns and $\mu_0 H_x =$ 100 mT.

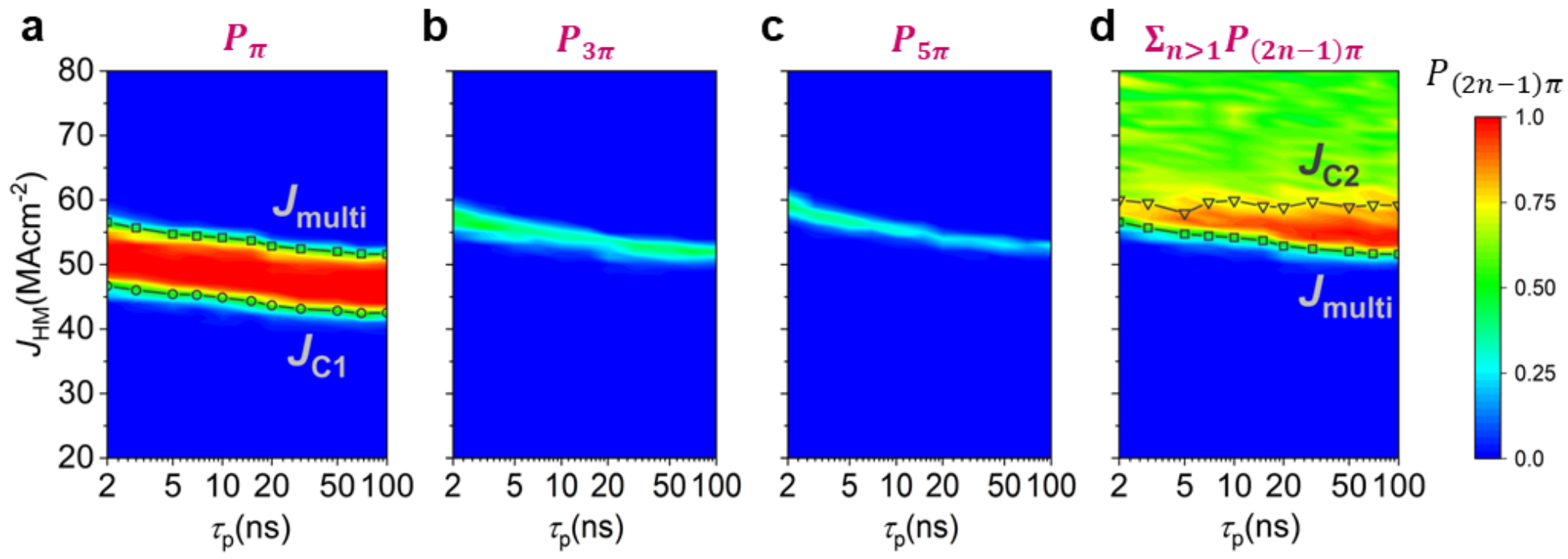

**Fig. 4. Numerical results for the probabilities of multiple rotations of the Néel vector.** **a-c**, Probabilities of the Néel vector completing $\pi$ (**a**), $3\pi$ (**b**) and $5\pi$ (**c**) rotations, as functions of $J_{\mathrm{HM}}$ and $\tau_P$. **d,** Summation of the probabilities of $3\pi$, $5\pi$, $7\pi$, … rotations, corresponding to the probability of multi-rotational switching.

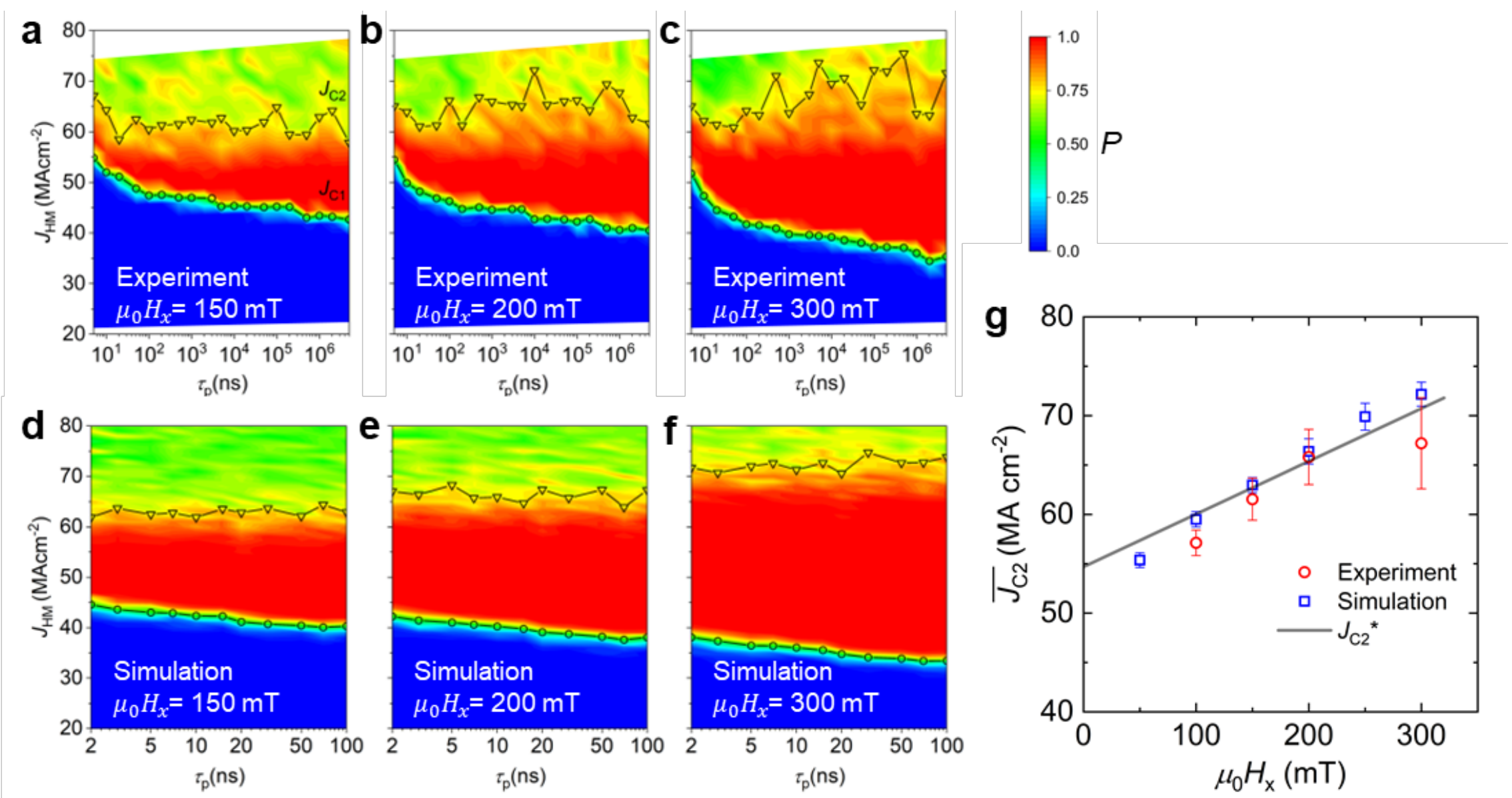

**Fig. 5. Magnetic-field dependence of the switching properties. a-f,** $P(J_{\mathrm{HM}}, \tau_P)$ for three different $H_x$, obtained by experiment (**a-c**) and simulation (**d-f**). **g,** $H_x$ dependence of the experimental and numerical $\overline{J_{\mathrm{C2}}}$ and the theoretical $J_{\mathrm{C2}}^{*}$. The error bars indicate the standard deviation.

## References

1. T. Jungwirth, X. Marti, P. Wadley, and J. Wunderlich, "Antiferromagnetic spintronics," *Nat. Nanotechnol.* **11**, 231–241 (2016).

2. V. Baltz *et al.*, "Antiferromagnetic spintronics," *Rev. Mod. Phys.* **90**, 015005 (2018).

3. J. Železný, P. Wadley, K. Olejník, A. Hoffmann, and H. Ohno, "Spin transport and spin torque in antiferromagnetic devices," *Nat. Phys.* **14**, 220–228 (2018).

4. J. Han, R. Cheng, L. Liu, H. Ohno, and S. Fukami, "Coherent antiferromagnetic spintronics," *Nat. Mater.* **22**, 684–695 (2023).

5. A. V. Kimel *et al.*, "Inertia-driven spin switching in antiferromagnets," *Nat. Phys.* **5**, 727–731 (2009).

6. K. Olejník *et al.*, "Terahertz electrical writing speed in an antiferromagnetic memory," *Sci. Adv.* **4**, eaar3566 (2018).

7. Y. Takeuchi *et al.*, "Electrical coherent driving of chiral antiferromagnet," *Science* **389**, 830–834 (2025).

8. H. Tsai *et al.*, "Picosecond ultralow-power switching device based on an antiferromagnet," *Science* **392**, 761–765 (2026).

9. H. Chen, Q. Niu, and A. H. MacDonald, "Anomalous Hall effect arising from noncollinear antiferromagnetism," *Phys. Rev. Lett.* **112**, 017205 (2014).

10. J. Kübler and C. Felser, "Non-collinear antiferromagnets and the anomalous Hall effect," *EPL* **108**, 67001 (2014).

11. S. Nakatsuji, N. Kiyohara, and T. Higo, "Large anomalous Hall effect in a non-collinear antiferromagnet at room temperature," *Nature* **527**, 212–215 (2015).

12. A. K. Nayak *et al.*, "Large anomalous Hall effect driven by a nonvanishing Berry curvature in the noncollinear antiferromagnet $Mn_3Ge$," *Sci. Adv.* **2**, e1501870 (2016).

13. X. Li *et al.*, "Anomalous Nernst and Righi-Leduc effects in $Mn_3Sn$: Berry curvature and entropy flow," *Phys. Rev. Lett.* **119**, 056601 (2017).

14. J. Liu and L. Balents, "Anomalous Hall effect and topological defects in antiferromagnetic Weyl semimetals: $Mn_3Sn$/Ge," *Phys. Rev. Lett.* **119**, 087202 (2017).

15. W. Feng, G.-Y. Guo, J. Zhou, Y. Yao, and Q. Niu, "Large magneto-optical Kerr effect

in noncollinear antiferromagnets $Mn_3X$ (X = Rh, Ir, Pt)," *Phys. Rev. B* **92**, 144426 (2015).

16. T. Higo *et al.*, "Large magneto-optical Kerr effect and imaging of magnetic octupole domains in an antiferromagnetic metal," *Nat. Photonics* **12**, 73–78 (2018).

17. M. Ikhlas *et al.*, "Large anomalous Nernst effect at room temperature in a chiral antiferromagnet," *Nat. Phys.* **13**, 1085–1090 (2017).

18. J. Železný *et al.*, "Spin-polarized current in noncollinear antiferromagnets," *Phys. Rev. Lett.* **119**, 187204 (2017).

19. M. Kimata *et al.*, "Magnetic and magnetic inverse spin Hall effects in a non-collinear antiferromagnet," *Nature* **565**, 627–630 (2019).

20. X. Chen *et al.*, "Octupole-driven magnetoresistance in an antiferromagnetic tunnel junction," *Nature* **613**, 490–495 (2023).

21. J. Han *et al.*, "Room-temperature flexible manipulation of the quantum-metric structure in a topological chiral antiferromagnet," *Nat. Phys.* **20**, 1110–1117 (2024).

22. Y. Takeuchi *et al.*, "Chiral-spin rotation of non-collinear antiferromagnet by spin–orbit torque," *Nat. Mater.* **20**, 1364–1370 (2021).

23. G. Q. Yan *et al.*, "Quantum sensing and imaging of spin–orbit-torque-driven spin dynamics in the non-collinear antiferromagnet $Mn_3Sn$," *Adv. Mater.* **34**, 2200327 (2022).

24. A. Manchon and S. Zhang, "Theory of nonequilibrium intrinsic spin torque in a single nanomagnet," *Phys. Rev. B* **78**, 212405 (2008).

25. I. M. Miron *et al.*, "Perpendicular switching of a single ferromagnetic layer induced by in-plane current injection," *Nature* **476**, 189–193 (2011).

26. L. Liu *et al.*, "Spin-torque switching with the giant spin Hall effect of tantalum," *Science* **336**, 555–558 (2012).

27. S. Fukami, T. Anekawa, C. Zhang, and H. Ohno, "A spin–orbit torque switching scheme with collinear magnetic easy axis and current configuration," *Nat. Nanotechnol.* **11**, 621–625 (2016).

28. E. V. Gomonay and V. M. Loktev, "Using generalized Landau-Lifshitz equations to describe the dynamics of multi-sublattice antiferromagnets induced by spin-polarized current," *Low Temp. Phys.* **41**, 698 (2015).

29. H. Fujita, "Field-free, spin-current control of magnetization in non-collinear chiral

antiferromagnets," *Phys. Status Solidi RRL* **11**, 1600360 (2017).

30. Y. Yamane, O. Gomonay, and J. Sinova, "Dynamics of noncollinear antiferromagnetic textures driven by spin current injection," *Phys. Rev. B* **100**, 054415 (2019).

31. H. Tsai *et al.*, "Electrical manipulation of a topological antiferromagnetic state," *Nature* **580**, 608–613 (2020).

32. T. Higo *et al.*, "Perpendicular full switching of chiral antiferromagnetic order by current," *Nature* **607**, 474–479 (2022).

33. B. Pal *et al.*, "Setting of the magnetic structure of chiral kagome antiferromagnets by a seeded spin-orbit torque," *Sci. Adv.* **8**, eabo5930 (2022).

34. G. K. Krishnaswamy *et al.*, "Time-dependent multistate switching of topological antiferromagnetic order in $Mn_3Sn$," *Phys. Rev. Appl.* **18**, 024064 (2022).

35. J.-Y. Yoon *et al.*, "Handedness anomaly in a non-collinear antiferromagnet under spin–orbit torque," *Nat. Mater.* **22**, 1106–1113 (2023).

36. Y. Kobayashi, Y. Shiota, H. Narita, T. Ono, and T. Moriyama, "Pulse-width dependence of spin–orbit torque switching in Mn3Sn/Pt thin films," *Appl. Phys. Lett.* **122**, 122405 (2023).

37. A. F. Andreev and V. I. Marchenko, "Symmetry and the macroscopic dynamics of magnetic materials," *Sov. Phys. Usp.* **23**, 21 (1980).

38. Z. Zhou *et al.*, "Field-free full switching of chiral antiferromagnetic order," Nature **651**, 341–347 (2026).

39. J. Yoon *et al.*, "Crystal orientation and anomalous Hall effect of sputter-deposited non-collinear antiferromagnetic Mn3Sn thin films," *Appl. Phys. Express* **13**, 013001 (2019).

40. Y. Sato *et al.*, "Thermal stability of non-collinear antiferromagnetic $Mn_3Sn$ nanodot," *Appl. Phys. Lett.* **122**, 122404 (2023).

41. A. Shukla, S. Qian, and S. Rakheja, "Impact of strain on the SOT-driven dynamics of thin film $Mn_3Sn$," *J. Appl. Phys.* **135**, 123903 (2024).

42. R. H. Koch, J. A. Katine, and J. Z. Sun, "Time-resolved reversal of spin-transfer switching in a nanomagnet," *Phys. Rev. Lett.* **92**, 088302 (2004).

43. E. M. Lifshitz and L. P. Pitaevskii, *Statistical Physics*, Course of Theoretical Physics

Vol. 9, Pt. 2 (Pergamon, Oxford, 1980).

44. W. F. Brown, “Thermal fluctuations of a single-domain particle,” *Phys. Rev.* **130**, 1677–1686 (1963).

45. M.-T. Suzuki, T. Koretsune, M. Ochi, and R. Arita, “Cluster multipole theory for anomalous Hall effect in antiferromagnets,” *Phys. Rev. B* **95**, 094406 (2017).

# Materials and methods

## Sample preparation

The MgO layer was deposited by RF sputtering at room temperature, while W, Ta, $Mn_3Sn$, and Ru layers were deposited by DC sputtering at 400 °C on a heated stage. The deposited stack was annealed at 500 °C for an hour in a vacuum. The buffer layer W(2 nm)/Ta(3 nm) is for epitaxial growth of the M-plane oriented hexagonal $D0_{19}$-$Mn_3Sn$[22,39]. The structural properties of our sample are similar to those in our previous study[7]. The stack was processed into a Hall device using electron beam lithography and Ar ion milling. To protect the channel layer, we deposited Ta(3 nm)/Ru(1 nm) films after the fabrication of the $Mn_3Sn$ nanodot. The coplanar waveguide and contact pads made of Cr(5 nm)/Au(100 nm) electrodes were formed after fabricating the channel.

## DC measurements

In the DC-regime measurements (Fig. **1c** and **d**), a current source (Yokogawa; Model GS200) was used for electric current application into the channel. The Hall voltage was measured using a voltmeter (Keithley; Model 2182A) with a DC current of 0.1 mA, corresponding to $J_{\mathrm{HM}} =$ 4.26 MA cm$^{-2}$. The shunt current was estimated by a method we reported in our previous study[7]. The $R_{\mathrm{H}}$-$H_z$ hysteresis curve was obtained by sweeping $H_z$ at a rate of 10 mT/s. For the $R_{\mathrm{H}}$-$J_{\mathrm{HM}}$ measurement, we injected a 100 ms-pulse from the current source.

## Switching probability measurement

In the switching probability measurements (Fig. **2a-e** and Fig. **5a-c**), pulse and DC voltages were applied into the channel from a function generator (Keysight; Model 81160). The Hall voltage was measured by a voltmeter (Keithley; Model 2182A) in the presence of a DC voltage of 30 mV, corresponding to $J_{\mathrm{HM}} =$ 3.42 MA cm$^{-2}$. The transmitted waveforms of the applied pulse and DC voltages were monitored using an oscilloscope (Agilent; Model DSOX93204A) connected to the output side of the waveguide. Supplementary Fig. 1 shows the measured waveforms of the applied voltage ($V_{\mathrm{OSC}}$) for four different $\tau_P$. The amplitude of the pulse current is determined by $V_{\mathrm{OSC}}/R_{\mathrm{OSC}}$, where $R_{\mathrm{OSC}}$ (= 50 Ω) is the internal impedance of the oscilloscope.

The procedure of the switching-probability measurement is schematically shown in Supplementary Fig. **2a**. First, the octupole state in the $Mn_3Sn$ nanodot was initialized to a low $V_{\mathrm{H}}$ state by a pulse with $J_{\mathrm{HM}} = -55$ MA cm$^{-2}$ and $\tau_P =$ 5 ms, under an in-plane magnetic field. The histogram of the measured $V_{\mathrm{H}}$ after the initialization is displayed in Supplementary Fig. **2b**, which confirms the successful initialization with no writing error. Next, the writing pulse was applied and then $V_{\mathrm{H}}$ was measured. This whole sequence was repeated 60 times to evaluate $P$ for given $(J_{\mathrm{HM}}, \tau_P)$. Supplementary Figs. **2c-f** show the distributions of $V_{\mathrm{H}}$ against $J_{\mathrm{HM}}$ for $\tau_{\mathrm{P}} =$ 5 ns, 500 ns, 50 $\mu$s and 5 ms, respectively, with $\mu_0 H_x$ = 200 mT. The gap between the high and low values of $V_{\mathrm{H}}$ is sufficiently large for an unambiguous evaluation of $P$.

## Numerical simulation

We model our $Mn_3Sn$ nanodot by the following magnetic energy density

$$u = J_0 \sum_{i,j} \boldsymbol{m}_i \cdot \boldsymbol{m}_j - D_0 \boldsymbol{e}_y \cdot \sum_{i,j} \boldsymbol{m}_i \times \boldsymbol{m}_j - K \sum_{i=A,B,C} \left(\boldsymbol{m}_i \cdot \boldsymbol{e}_{K,i}\right)^2 - \frac{K_z}{3} \sum_{i=A,B,C} \left(\boldsymbol{m}_i \cdot \boldsymbol{e}_{Kz,i}\right)^2 - M_s \boldsymbol{H} \cdot \sum_{i=A,B,C} \boldsymbol{m}_i \quad (3),$$

where $J_0$ is the antiferromagnetic exchange coupling constant, $D_0$ is the Dzyaloshinskii-Moriya coupling constant, $K$ is the uniaxial anisotropy constant originating from the local crystalline symmetry, $K_z$ represents the global uniaxial anisotropy induced predominantly by the interface, and $\boldsymbol{H}$ is the external magnetic field. The unit vectors $\boldsymbol{e}_{K,i}$ and $\boldsymbol{e}_{Kz,i}$, representing the easy axes of the anisotropies, are given by $\boldsymbol{e}_{K,\mathrm{A}} = -\frac{1}{2}\boldsymbol{e}_x + \frac{\sqrt{3}}{2}\boldsymbol{e}_z$, $\boldsymbol{e}_{K,\mathrm{B}} = -\frac{1}{2}\boldsymbol{e}_x - \frac{\sqrt{3}}{2}\boldsymbol{e}_z$, $\boldsymbol{e}_{K,\mathrm{C}} = \boldsymbol{e}_x$, $\boldsymbol{e}_{Kz,\mathrm{A}} = \frac{\sqrt{3}}{2}\boldsymbol{e}_x - \frac{1}{2}\boldsymbol{e}_z$, $\boldsymbol{e}_{Kz,\mathrm{B}} = -\frac{\sqrt{3}}{2}\boldsymbol{e}_x - \frac{1}{2}\boldsymbol{e}_z$, and $\boldsymbol{e}_{Kz,\mathrm{C}} = \boldsymbol{e}_z$. Here we set the $x$, $y$, and $z$ axes along $[11\bar{2}0]$, $[0001]$ and $[1\bar{1}00]$, respectively, of $Mn_3Sn$.

The coupled LLG equations for the sublattice magnetizations are given by

$$\frac{\partial \boldsymbol{m}_i}{\partial t} = -\gamma \boldsymbol{m}_i \times \left( \boldsymbol{H}_{\mathrm{eff}}^i + \boldsymbol{H}_{\mathrm{th}}^i \right) + \alpha \boldsymbol{m}_i \times \frac{\partial \boldsymbol{m}_i}{\partial t} - \gamma D_{\mathrm{s}} J_{\mathrm{HM}} \boldsymbol{m}_i \times (\boldsymbol{m}_i \times \boldsymbol{s}) \quad (4),$$

where $\boldsymbol{H}_{\mathrm{eff}}^i = -\frac{1}{M_s}\frac{\partial u}{\partial \boldsymbol{m}_i}$ is the effective magnetic field exerted on $\boldsymbol{m}_i$, and $\boldsymbol{H}_{\mathrm{th}}^i$ describes thermal fluctuations satisfying $\langle H_{\mathrm{th}}^{i,\mu}(t) \rangle = 0$ and $\langle H_{\mathrm{th}}^{i,\mu}(t)\, H_{\mathrm{th}}^{i,\nu}(t') \rangle = \frac{2\alpha k_{\mathrm{B}} T}{\gamma M_s V} \delta_{\mu\nu} \delta(t - t')$, with $\mu, \nu = x, y, z$, $k_{\mathrm{B}}$ the Boltzmann constant, $V = \frac{\pi D^2 d}{4}$ the volume of the $Mn_3Sn$ nanodot, and $D$ and $d$ the diameter and thickness of the nanodot, respectively. The last term in Eq. (4) represents the SOT with $D_{\mathrm{s}} = \frac{1}{3}\frac{\hbar \theta_{\mathrm{SH}}}{2eM_s d}$ and $\theta_{\mathrm{SH}}$ denoting the effective spin Hall angle. The factor 1/3 in $D_{\mathrm{s}}$ reflects the sublattice symmetry where the spin current injected into $Mn_3Sn$ is transferred equiprobably to each of the three magnetic sublattices.

The antiferromagnetic order parameters for a noncollinear AFM can be constructed as[28,30]

$$\boldsymbol{n}_1 = \frac{\boldsymbol{m}_{\mathrm{A}} + \boldsymbol{m}_{\mathrm{B}} - 2\boldsymbol{m}_{\mathrm{C}}}{3\sqrt{2}}, \qquad \boldsymbol{n}_2 = \frac{-\boldsymbol{m}_{\mathrm{A}} + \boldsymbol{m}_{\mathrm{B}}}{\sqrt{6}}. \tag{5}$$

Assuming the net magnetization to be sufficiently small as $|\boldsymbol{m}| = \frac{|\boldsymbol{m}_{\mathrm{A}}+\boldsymbol{m}_{\mathrm{B}}+\boldsymbol{m}_{\mathrm{C}}|}{3} \ll 1$, it is straightforward to show $\boldsymbol{n}_1 \cdot \boldsymbol{n}_2 \simeq 0$ and $|\boldsymbol{n}_1| \simeq |\boldsymbol{n}_2| \simeq 1/\sqrt{2}$. Therefore, tracking either one of $\boldsymbol{n}_1$ or $\boldsymbol{n}_2$ is sufficient to capture the dynamics of the noncollinear AFM order in $Mn_3Sn$. $\boldsymbol{n}_1$ can be interpreted as corresponding to the octupole moment, and in the main text, we adopted the notation $\boldsymbol{n} \equiv \boldsymbol{n}_1$.

For the simulations discussed in the main text, we employed the following parameters: $J_0 = 8 \times 10^7$ Jm$^{-3}$, $D_0 = 6 \times 10^6$ Jm$^{-3}$, $K = 2.5 \times 10^5$ Jm$^{-3}$, $K_z = 9 \times 10^2$ Jm$^{-3}$, $M_{\mathrm{s}} = 0.5$ T, $\gamma = 2.21 \times 10^5$ mA$^{-1}$s$^{-1}$, $\alpha = 0.005$, $\theta_{\mathrm{SH}} = -0.1$, $d = 20$ nm, $D = 200$ nm, and $T = 300$ K.

## Theoretical analysis

Our derivation of Eq. (1) in the main text follows Ref. [30]. Here we outline the derivation briefly. We first rewrite the LLG equations (4) in terms of $(\boldsymbol{n}_1, \boldsymbol{n}_2, \boldsymbol{m})$, and introduce the angular variable by $\boldsymbol{n}_1 = \frac{1}{\sqrt{2}}(\sin\theta, 0, \cos\theta)$. Because of the conditions $\boldsymbol{n}_1 \cdot \boldsymbol{n}_2 \simeq 0$ and $D_0 > 0$, one finds $\boldsymbol{n}_2 = \frac{1}{\sqrt{2}}\left[\sin\left(\theta + \frac{\pi}{2}\right), 0, \cos\left(\theta + \frac{\pi}{2}\right)\right]$. Taking advantage of the condition $|\boldsymbol{m}| \ll 1$, a perturbative treatment of $\boldsymbol{m}$ up to its first order leads to an expression of $\boldsymbol{m}$ as a function of $\boldsymbol{n}_1$ [28,30]. Putting all these together, we arrive at the equation of motion for $\theta$, that is Eq. (1), where $H_\mathrm{E} = \frac{3J_0}{M_s}$ and $H_K = \frac{2K_z}{3M_s}$. The uncompensated net magnetization is given by $M_\mathrm{UC} = \frac{M_\mathrm{S} K}{J_0}$ in equilibrium without an external field.

$J_{\mathrm{C}1}^*$ and $J_{\mathrm{C}2}^*$ are derived from Eq. (2) as follows. A critical current density and the corresponding critical angle $\theta^*$ are defined as $(J_\mathrm{C}, \theta)$ that simultaneously satisfy $\partial_\theta u = 0$ and $\partial_\theta^2 u = 0$. Assuming that $|H_x|$ is sufficiently small that it hardly affects $\theta^*$, we find $\theta^* \cong \frac{\pi}{4}, \frac{3\pi}{4}, \frac{5\pi}{4}, \frac{7\pi}{4}$. The critical angle corresponding to the down-to-up transition, with the initial state $\theta = \pi$, is therefore $\frac{5\pi}{4}$. Substituting $\theta = \frac{5\pi}{4}$ to $\partial_\theta u = 0$, we reach the expression of $J_{\mathrm{C}1}^*$ given in the main text. $J_{\mathrm{C}2}^*$ can be obtained in a similar fashion considering the up-to-down transition.

## Switching probability at zero field

Supplementary Fig. **3** shows the experimental $P(J_{\mathrm{HM}}, \tau_P)$ measured in the absence of external magnetic field. There appears only one critical current density that separates the non-switched (blue color) and the coherent rotation (green color) regimes. The absence of the switched regime can also be derived analytically by setting $H_x = 0$ in Eq. (2).

## Nanodot-diameter dependence of the switching probability

Supplementary Fig. **4** shows the calculated $P(J_{\mathrm{HM}}, \tau_P)$ with $\mu_0 H_x = 100$ mT for different nanodot diameters $D$, corresponding to different levels of thermal fluctuation (see the definition of the thermal fields $\boldsymbol{H}_{\mathrm{th}}^{i}$ given above). As $D$ decreases, i.e., as thermal fluctuations increase, $J_{\mathrm{C1}}$ becomes lower. This is because the $\pi$ switching with thermal assistance takes place more readily under stronger thermal fluctuations. In contrast, $J_{\mathrm{C2}}$ is largely insensitive to thermal fluctuations, consistent with the physical picture discussed in the main text.

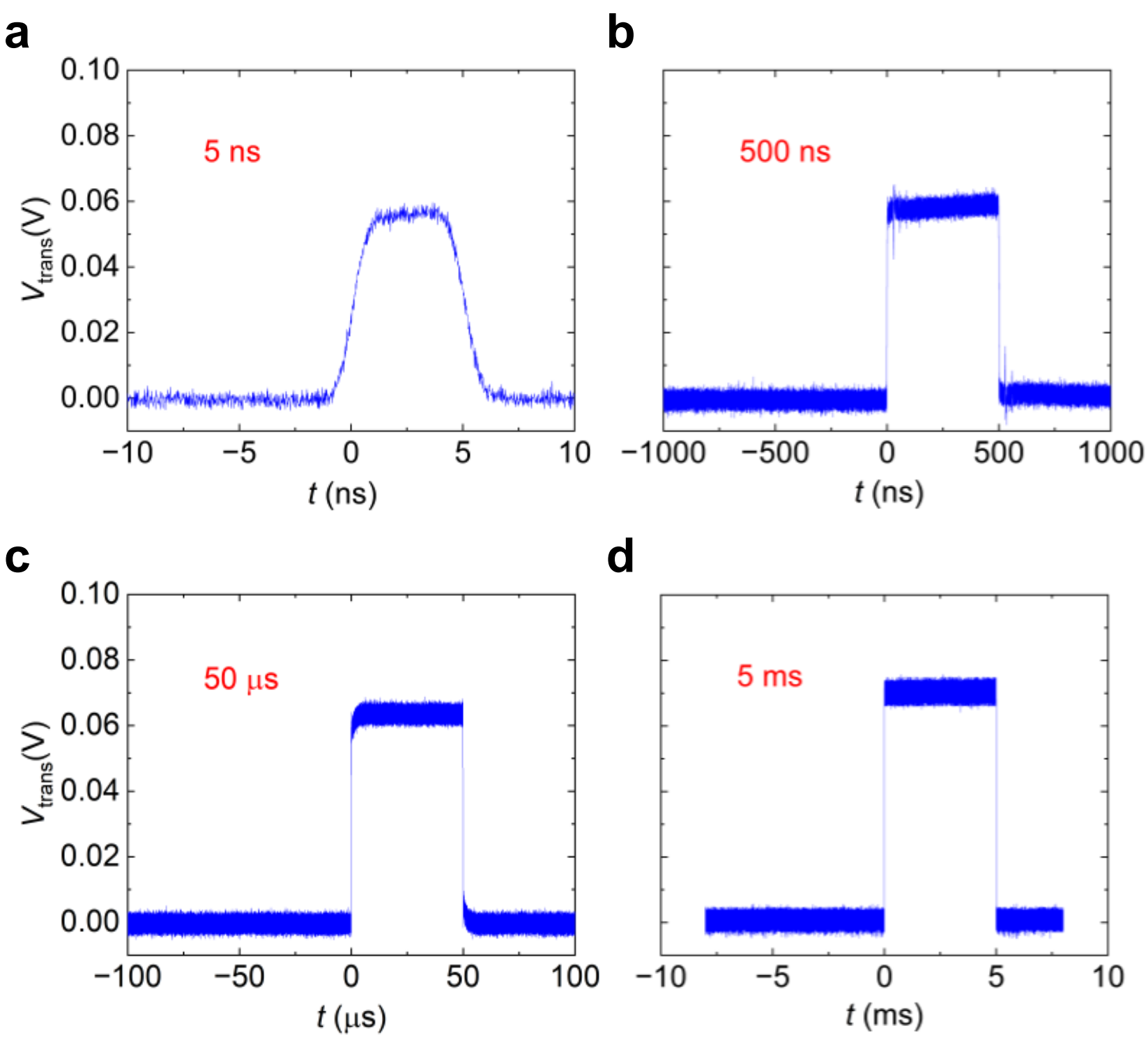

**Supplementary Fig. 1. Transmitted waveforms of the applied pulse voltages.** The transmitted voltage $V_{\mathrm{trans}}$ with an input pulse voltage of 500 mV is plotted for four different $\tau_P$.

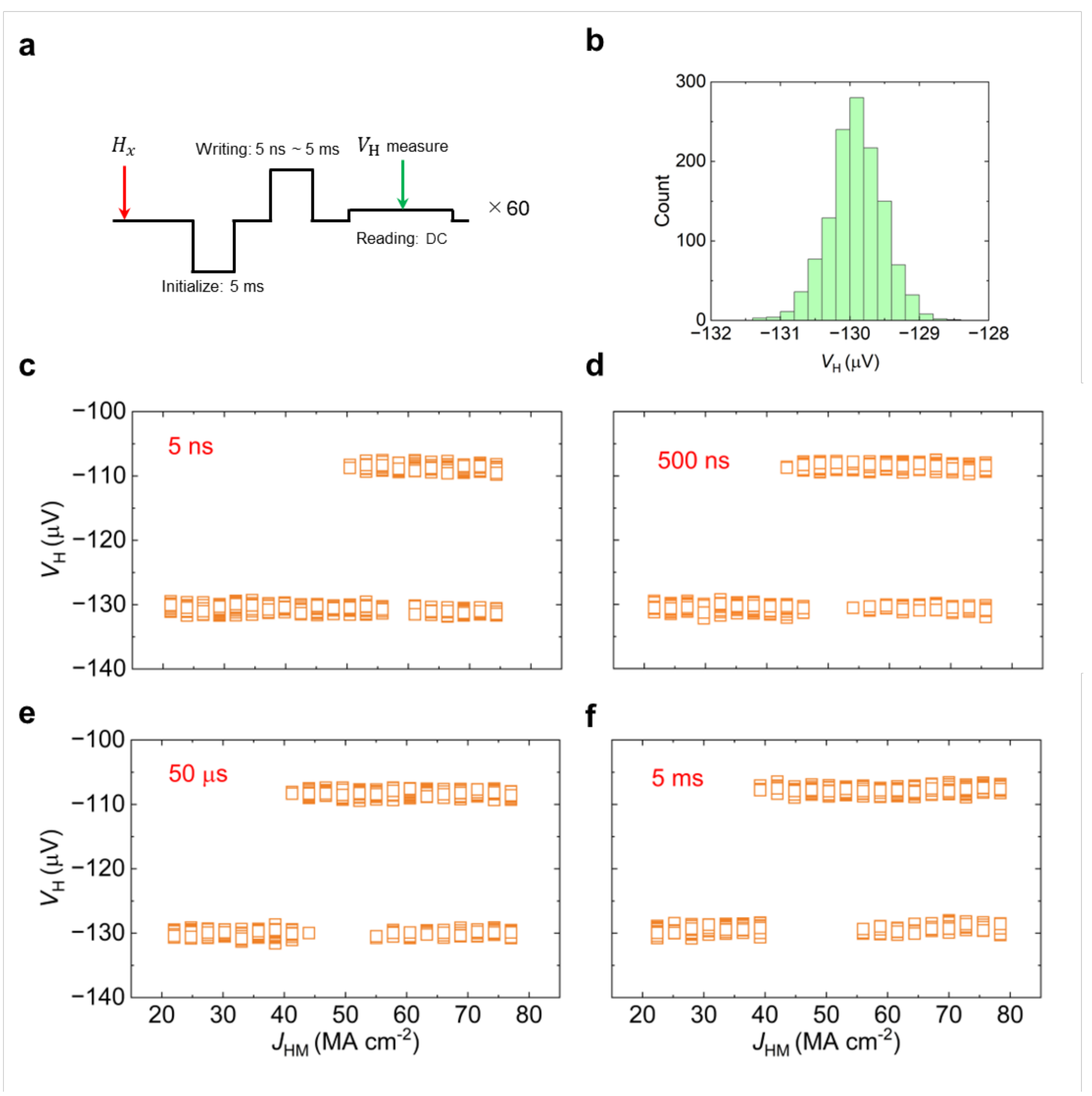

**Supplementary Fig. 2. Scheme for the switching probability measurement. a**, Procedure of the switching probability measurement. **b**, Histogram of $V_H$ measured after the initialization. **c-f**, Distribution of $V_H$ against $J_{HM}$, measured after applications of the writing pulse with four different $\tau_P$.

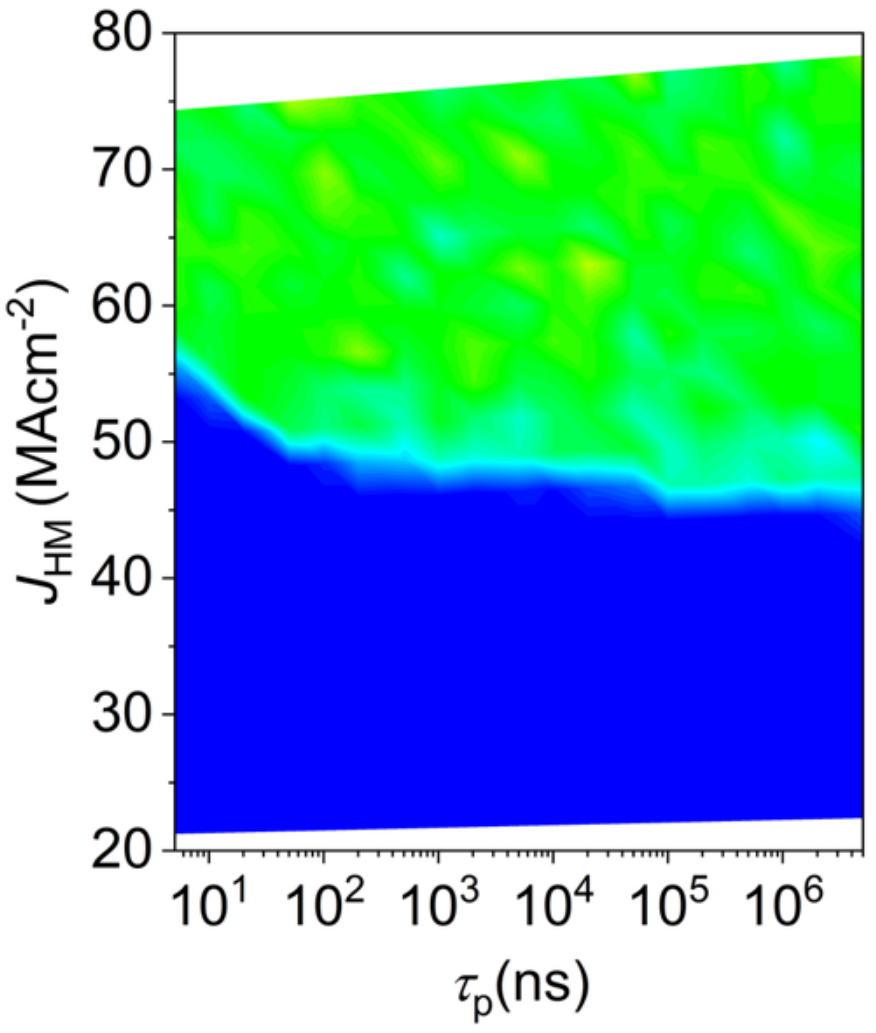

**Supplementary Fig. 3. Switching probability with no external field.** The experimentally observed $P(J_{\mathrm{HM}}, \tau_P)$ with $H_x = 0$.

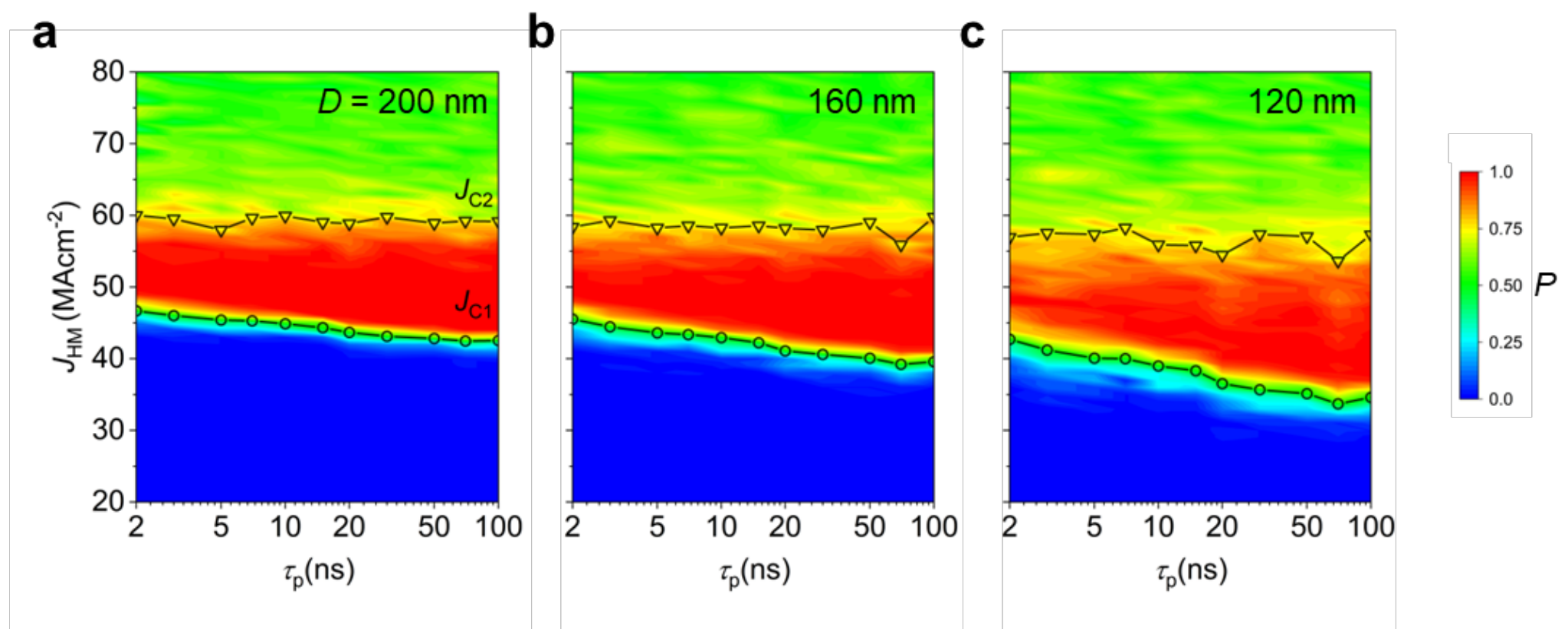

**Supplementary Fig. 4. Numerical results for the nanodot-diameter dependence of the switching probability.** $P(J_{\mathrm{HM}}, \tau_P)$ with $\mu_0 H_x = 100$ mT for three different nanodot diameters $D$ are plotted: $D = 200$ nm (**a**), 160 nm (**b**), and 120 nm (**c**).

## Acknowledgements

The authors thank J.-Y. Yoon and K. V. D. Zoysa for valuable discussions, and Y. Nakano, T. Tanno, I. Morita, R. Ono, and M. Musya for their technical support. This work was supported by Japan Society for the Promotion of Science KAKENHI (grant nos. 23K26521, 24H00039, 24H02235, 24KJ0432 and 25K01276), Japan Science and Technology Agency PRESTO (grant no. JPMJPR24H6) and TI-FRIS program, Iketani Science and Technology Foundation, and Seiko Instruments Advanced Technology Foundation.